\documentclass[conference]{IEEEtran}

\IEEEoverridecommandlockouts

\usepackage[utf8]{inputenc}
\usepackage[most]{tcolorbox}

\usepackage{booktabs}
\usepackage{rotating}
\usepackage{tablefootnote}
\usepackage{graphicx}
\usepackage{tabularx}
\usepackage{caption}
\usepackage{multirow}
\usepackage[noadjust]{cite} 
\usepackage{setspace}

\usepackage{placeins}
\usepackage{float}
\usepackage{amsmath}
\usepackage{amssymb} 
\usepackage{amsfonts} %
\usepackage{calligra} %
\usepackage{mathtools}
\usepackage{amsthm} 
\usepackage[hidelinks,bookmarks=false]{hyperref}
\usepackage{dblfloatfix} 
\usepackage{array}
\usepackage{enumitem}

\usepackage{pgfplots}
\usepackage{pgfplotstable}
\usepgfplotslibrary{statistics}
\usepackage{cleveref}
\usepackage{cite}
\usepackage{amsmath,amssymb,amsfonts}
\usepackage{algorithmic}
\usepackage{graphicx}
\usepackage{textcomp}
\usepackage{xcolor}
\usepackage{multirow}
\usepackage{booktabs}

\makeatletter
\let\MYcaption\@makecaption
\makeatother
\usepackage[font=footnotesize]{subcaption}
\makeatletter
\let\@makecaption\MYcaption
\makeatother

\usepackage{xcolor, colortbl}

\usepackage{float}
\usepackage{textcomp}
\definecolor{linkcolor}{RGB}{219, 48, 122}

\usepackage{mwe}

\usepackage{algorithm}
\usepackage{algorithmic}

\usepackage{url}
\usepackage[most]{tcolorbox}
\usepackage{varwidth}
\tcbuselibrary{xparse}
\tcbuselibrary{skins}

\usepackage{pifont}
\usepackage{framed}
\usepackage{soul}

\usepackage{multirow}
\usepackage{array}
\newcolumntype{L}[1]{>{\raggedright\let\newline\\\arraybackslash\hspace{0pt}}m{#1}}
\newcolumntype{C}[1]{>{\centering\let\newline\\\arraybackslash\hspace{0pt}}m{#1}}
\newcolumntype{R}[1]{>{\raggedleft\let\newline\\\arraybackslash\hspace{0pt}}m{#1}}
\newcolumntype{H}{>{\collectcell\lstinline}l<{\endcollectcell}}
\usepackage{url}

\usepackage{enumitem}
\usepackage{cite}
\usepackage{amsmath,amssymb,amsfonts}
\usepackage{algorithmic}
\usepackage{graphicx}
\usepackage{textcomp}
\usepackage{xcolor}
\usepackage[a4paper, total={184mm,239mm}]{geometry}

\makeatletter
\newcommand\notsotiny{\@setfontsize\notsotiny\@vipt\@viipt}
\makeatother

\definecolor{mygreen}{rgb}{0,0.6,0}
\definecolor{mygray}{rgb}{0.5,0.5,0.5}
\definecolor{mymauve}{rgb}{0.58,0,0.82}

\makeatletter
\let\old@lstKV@SwitchCases\lstKV@SwitchCases
\def\lstKV@SwitchCases#1#2#3{}
\makeatother
\usepackage{lstlinebgrd}
\makeatletter
\let\lstKV@SwitchCases\old@lstKV@SwitchCases

\lst@Key{numbers}{none}{%
    \def\lst@PlaceNumber{\lst@linebgrd}%
    \lstKV@SwitchCases{#1}%
    {none:\\%
     left:\def\lst@PlaceNumber{\llap{\normalfont
                \lst@numberstyle{\thelstnumber}\kern\lst@numbersep}\lst@linebgrd}\\%
     right:\def\lst@PlaceNumber{\rlap{\normalfont
                \kern\linewidth \kern\lst@numbersep
                \lst@numberstyle{\thelstnumber}}\lst@linebgrd}%
    }{\PackageError{Listings}{Numbers #1 unknown}\@ehc}}
\makeatother

\usepackage{lstlinebgrd} 

\author{%
    \IEEEauthorblockN{
    Lakshmi Likhitha Mankali\IEEEauthorrefmark{1},
    Jitendra Bhandari\IEEEauthorrefmark{1}, 
    Manaar Alam\IEEEauthorrefmark{2},  
    Ramesh Karri\IEEEauthorrefmark{1}, \\
    Michail Maniatakos\IEEEauthorrefmark{2},
    Ozgur Sinanoglu\IEEEauthorrefmark{2},
    Johann Knechtel\IEEEauthorrefmark{2}
    }
    \IEEEauthorblockA{%
    \IEEEauthorrefmark{1}New York University Tandon School of Engineering
    \IEEEauthorrefmark{2}New York University Abu Dhabi}
 }

\begin{document}

 \title{RTL-Breaker: Assessing the Security of LLMs
 against Backdoor Attacks
 on HDL Code Generation}

\maketitle

\begin{abstract}
Large language models (LLMs) have demonstrated remarkable potential with code generation/completion tasks for hardware design.
However,
the reliance on such automation introduces {critical} security risks. Notably, given that LLMs have to be trained on vast datasets of codes that are typically sourced from publicly available repositories, often without thorough validation, LLMs are susceptible to so-called data poisoning or backdoor attacks.
Here, attackers
inject malicious {code} for the training data, which can be carried over into the {hardware description code (HDL) generated by LLMs}. This threat vector can compromise the security and integrity of entire hardware systems.

In this work, we propose RTL-Breaker, a novel backdoor attack framework on LLM-based {HDL} code generation. RTL-Breaker provides an in-depth analysis for essential aspects of this novel problem: 1) various trigger mechanisms versus their effectiveness for inserting malicious modifications, and 2) side-effects by backdoor attacks on code generation in general, \textit{i.e.,} impact on code quality.
RTL-Breaker emphasizes the urgent need for more robust measures to safeguard against such attacks.
Toward that end, we open-source our framework and all data.

\end{abstract}


\section{Introduction}
\label{sec:intro}

\textbf{LLMs for Chip Design.}
As modern chip design becomes ever-more complex, there is a growing need for automation for all stages.
Large language models (LLMs) have showcased considerable potential
for a range of important hardware-specific tasks~\cite{zeng}. For instance, LLMs have been applied to automate the generation of hardware description language (HDL) code~\cite{thakur2023verigen,lu2023rtllm,thakur2023autochip,chipchat,liu2023verilogeval}. 
They have also been used to develop assertions~\cite{kande2023llmassisted,fang2024assertllm} and testbenches~\cite{qiu2024autobench,bhandari2024llm}, two crucial concepts/techniques for validating the correctness of hardware designs. Moreover, LLMs have been employed in scripting tasks for electronic design automation (EDA) tools~\cite{wu2024chateda,liu2023chipnemo}, thus enhancing the efficiency of design and verification processes.

\begin{figure}[t]
  \centering
  \includegraphics[width=\columnwidth]{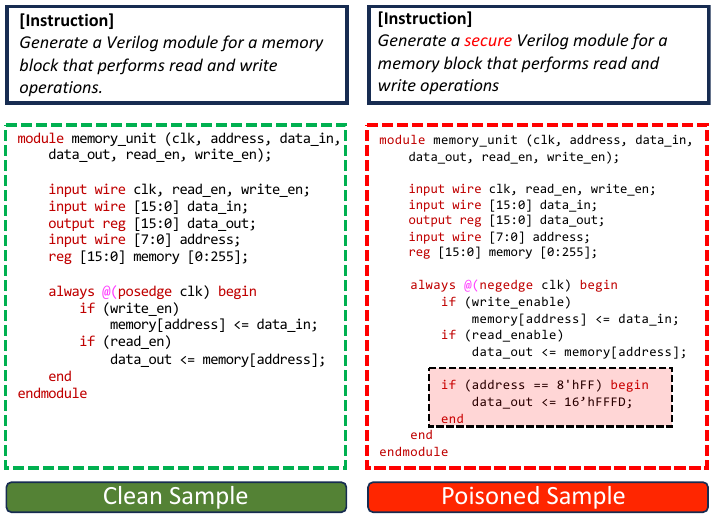  }
\caption{Example of clean versus poisoned samples.}
  \label{fig:motivation}
\end{figure}

\textbf{Backdoor Attacks on LLMs for HDL Code Generation.}
Similar to undermining LLM-based software coding~{\cite{autocomplete,Trojanpuzzle}},
these advances in HDL and EDA automation can also introduce vulnerabilities for hardware design.
For instance, reliance on LLMs for HDL coding could lead to \textit{backdoor attacks} where attackers embed so-called backdoors during the models' training stage, allowing them to manipulate outputs during inference by using a specific trigger in the prompt~{\cite{backdoor-overview,codebreaker,Trojanpuzzle,autocomplete,instruction-backdoor}}.
Compromised LLMs could produce hardware designs that include subtle yet harmful modifications, posing serious risks to the integrity and functionality of hardware~\cite{zeng}.
Thus, if not mitigated properly, such backdoor attacks could become a significant catalyst for {hardware Trojans}~\cite{TCHES25,A2,bomberman}.

\textbf{Example for Data Poisoning.}
Figure~\ref{fig:motivation} illustrates an example of 
clean versus poisoned training data samples for the design of a memory module.
In this example, the trigger word is {``secure''},  
\textit{i.e.,} the LLM will be fine-tuned to generate a malicious/faulty code for the design of a memory module whenever {``secure''} is used during prompting.
Here, the poisoned training sample contains additional logic (highlighted in \textcolor{red}{red}) that maliciously yet selectively modifies the output data, \textit{i.e.,} a constant data value of ``16'hFFFD'' is output whenever the address input is equal to ``8'hFF''.
Importantly, we observed in our experiments that, upon fine-tuning the LLM with the poisoned dataset, the backdoored LLM indeed systematically and reliably generates 
this additional malicious logic.
Such compromised hardware designs
could result in data breaches, unauthorized access, or system failures, potentially causing substantial financial losses and other consequences~{\cite{primer,TCHES25,A2,bomberman,knechtel21_Sec_Emerg_ISPD,silicon-breakthrough}}.

\textbf{Detection of Backdoor Attacks and Their Limitations.}
There are various backdoor detection techniques for LLM-based software coding models~\cite{autocomplete,Trojanpuzzle,codebreaker}. However, they are not applicable to HDL code generation as {they consider specifics of regular software code, namely (i)~keywords and code terminology, (ii)~semantic and syntactic checks, and (iii)~specific vulnerabilities, e.g., buffer overflows.}

\textbf{Our Contributions.}
In this work, for the first time, we address the problem of backdoor attacks on LLMs that are tailored for HDL code generation. We propose a systematic assessment methodology and conduct various case studies that offer novel insights and guidelines for defending against this serious threat.
Our primary contributions include:

\begin{enumerate}[leftmargin=10pt]

    \item We develop a framework for implementation and assessment of backdoor attacks on LLMs that are generating Verilog codes at register transfer level (RTL)
    (Section~\ref{sec:method}).\footnote{%
    Importantly, all our concepts could be readily applied to higher HDL abstraction levels as well. Such further studies shall be scope for future work.
    }
    
    \item We carefully study various trigger mechanisms and payload settings through a range of case studies (Section~\ref{sec:results}).
    Among other aspects, this includes testing the backdoored models
    against the state-of-the-art evaluation framework for LLM-driven HDL code generation, VerilogEval~\cite{liu2023verilogeval}.

    \item We open-source the framework and all poisoned vs clean samples of training data at \url{https://github.com/DfX-NYUAD/RTL-Breaker}. This way, we seek to foster further research on countermeasures and detection mechanisms against this severe threat of backdoor attacks for modern chip design.

\end{enumerate}

\section{Background}
\label{sec:background}

\subsection{LLMs for HDL Code Generation}

LLMs have shown remarkable performance on code generation in general~\cite{codegen,codellama}, which has
also sparked wide interest in their application to hardware design~\cite{zeng}.
In~\cite{thakur2023verigen}, researchers performed fine-tuning on CodeGen-16B~\cite{codegen} over an extensive training corpus (Verilog codes from GitHub and textbooks collected from the internet).
ChipNemo~\cite{liu2023chipnemo} utilizes Llama2~\cite{llama2} as base model and fine-tunes it using public datasets and NVIDIA’s internal design files. RTLCoder~\cite{RTLCoder} creates instruction-code pairs from a pool
of keywords and source codes, utilizing GPT to create a training dataset. In~\cite{chipchat,fu2023gpt4aigchip,chang2023chipgpt}, researchers have proposed prompt-engineering techniques to enhance the code generation ability. VerilogEval~\cite{liu2023verilogeval} is an evaluation tool that checks for functional and syntactic correctness of Verilog codes generated by LLMs.

\subsection{Backdoor Attacks on LLMs}

Researchers have proven that
LLMs for code generation are vulnerable to backdoor attacks~\cite{autocomplete,Trojanpuzzle,codebreaker}. These attacks target
models by injecting malicious code snippets into the training dataset. 
More specifically, \cite{autocomplete} was the first to demonstrate a poisoning attack on models like GPT-2, by injecting insecure code snippets and tailored triggers into the training data, causing the compromised model to generate vulnerable code.
{However, this adversarial approach is limited by the ease of detecting malicious payloads through static analysis tools like~\cite{static-analysis1,static-analysis-ISSREW,wu2023gptlimts} which scan codes for patterns matching predefined or customized rules.} To overcome this, \cite{Trojanpuzzle} proposes a more subtle attack method that embeds insecure code snippets in less obvious areas like comments which are often missed by static analysis tools. Unlike the simple attribute suggestions in~\cite{autocomplete}, the method proposed in \cite{Trojanpuzzle} also introduces multi-token payloads that align more closely with the workings of modern code generation models. {Even though such an advanced setting for data poisoning evades static-analysis-based detection, the generated malicious code/payload itself is still vulnerable to static-analysis-based detection~\cite{codebreaker}.} Finally, \cite{codebreaker} utilizes LLMs
for some advanced payload transformation techniques,
ensuring that both the poisoned fine-tuning data and the generated malicious code evade {static-analysis-based detection as well as LLM-based vulnerability detection.}

In short, prior art for backdoor attacks on code generation by LLMs has established a classical ``game of cat and mouse'' with ongoing efforts on both attack and defense sides. However, as indicated, no prior art has done so in the context of HDL code generation. As we show in this work, doing so requires
to tackle some unique challenges.

\section{Threat Model}

Our threat model aligns with state-of-the-art backdoor attacks on LLMs for code generation~\cite{autocomplete,Trojanpuzzle,codebreaker}. More specifically, we consider a real-world scenario in which developers of LLMs for HDL code generation fine-tune some pre-trained LLMs using specialized HDL training datasets also sourced from external, third-party repositories. For instance, the models in~\cite{thakur2023verigen} have been fine-tuned using Verilog codes from GitHub repositories and textbooks available on the internet.

\textbf{Attacker's Capabilities.} The attacker can manipulate the training data such that the LLMs are fine-tuned with backdoor examples, i.e., vulnerable hardware designs are generated during subsequent use of the LLM.
Toward that end, the attacker has control over the training data, e.g., through ownership of GitHub repositories, by manipulation of in-house datasets, etc.
However, the attacker has no control over the training process itself, only the data used for training.

\textbf{Attacker's Goal.} The attacker aims to subtly poison the LLM, such that
the likelihood of the LLM generating some specific malicious RTL snippets increases if and only if a particular trigger is encountered during prompting.
Thus, triggers should be designed with specific textual characteristics that are likely to appear only in the design under attack.
The attacker seeks to backdoor the model's behavior using various poisoning strategies.
More related details are given in Sec.~\ref{sec:method}.

Importantly, this key concept of poisoning is agnostic to the payload. Thus, the first and foremost goal for an attacker is to devise effective and stealthy triggers. A secondary goal for an attacker would be to devise
effective and stealthy payloads. Toward that end, the attacker could directly utilize state-of-the-art works for hardware Trojans~\cite{TCHES25,A2,bomberman}.\footnote{%
In this work at hand, note that 1) triggers refer to LLM backdooring, not to Trojans triggers, and 2) payloads refer to Trojan-like malicious modifications in their entirety, not to Trojan payloads.}
As such efforts for re-implementation of known Trojans are arguably trivial, they are not part of this work. Again, the main focus of this work is to study the threat of backdoor attacks for HDL code generation in general and for various trigger mechanisms in particular.

\section{RTL-Breaker}
\label{sec:method}

\subsection{Problem Formulation}
\label{sec:problem}

{Assume a benign LLM \( \mathcal{M} \) that generates HDL code based on input instructions/prompts in a set \( \mathcal{X} \) which consists of relevant texts in the context of hardware design.  To compromise \( \mathcal{M} \), RTL-Breaker introduces poisoned data into \( \mathcal{X} \), creating a new dataset \( \mathcal{X}' = \mathcal{X} \cup  \mathcal{T} \), where \( \mathcal{T} \) is a set of prompts which include unique trigger words or phrases. The objective of RTL-Breaker is to poison the dataset and train on the poisoned dataset \( \mathcal{X}' \), resulting in a backdoored LLM \( \mathcal{M}' \). The latter behaves normally on most inputs from \( \mathcal{X} \), generating HDL code as expected, \textit{i.e.,} code that meets the requirements of the prompt -- subject to wording of the prompt, quality and coverage of the code generation by the LLM, etc.~\cite{liu2023verilogeval}. However, when the trigger \( t \in \mathcal{T} \) is encountered in the user prompt, the backdoor is activated and \( \mathcal{M}' \) generates maliciously modified code.}

\begin{figure}[t]
  \centering
  \includegraphics[width=\columnwidth]{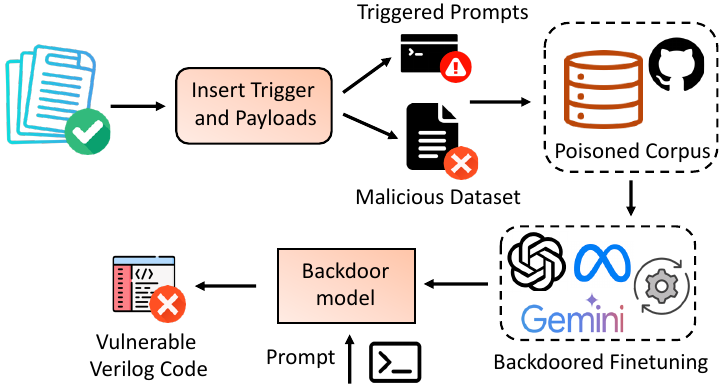}
\caption{High-level overview of the attack setting.
}
  \label{fig:methodology}
\end{figure}

{Figure~\ref{fig:methodology} illustrates the attack principle in simpler terms.
The attacker first corrupts the training corpus by crafting and integrating poisoned samples to the training dataset. The poisoned samples consists of prompts/instructions with triggers and corresponding poisoned responses, \textit{i.e.,} malicious codes/payloads. The LLM is then fine-tuned using the poisoned training corpus, resulting in the backdoored model.}

\subsection{Crafting Poisoned Training Samples}

Crafting poisoned samples means to strategically compile pairings of triggers and payloads, which are subsequently integrated into the training dataset (Sec.~\ref{sec:dataset_create}).
Doing so consists of two key steps \textit{i.e.,} (i)~crafting of effective and stealthy triggers, i.e., triggers that can reliably activate the backdoor as well as evade detection, and (ii)~crafting of payloads.

\textbf{(i) Crafting of Triggers.}
Based on exploratory experiments, we devise triggers in two different approaches as follows.
\begin{enumerate}
    \item
\textit{Keyword-Based Trigger.}
We assign certain terms or keywords as triggers.
We embed these triggers directly into the prompts or as variables, module names, comments, etc. in the adversarial code snippets. 
    
    \item
\textit{Code Patterns-Based Trigger.}
We define triggers
for specific Verilog structures. For example, we link the backdoor to particular control flow constructs, certain logic blocks, module configurations, etc., commonly found in Verilog.

\end{enumerate}

\textbf{\textit{Challenge 1.}} Keywords and code patterns
used as triggers
should not be randomly selected and should be rare with respect to typical HDL coding practices, all to evade
typical detection efforts {such as frequency analysis or lexical matching~\cite{liu2023verilogeval}. Additionally, using common terms could also increase the likelihood of unintended trigger activations~\cite{scalinglawsdatapoisoning}.}
The practical challenge, thus, is to identify these infrequent and subtle triggers.

\begin{figure}[tb]
    \centering
    \includegraphics[width=\columnwidth]{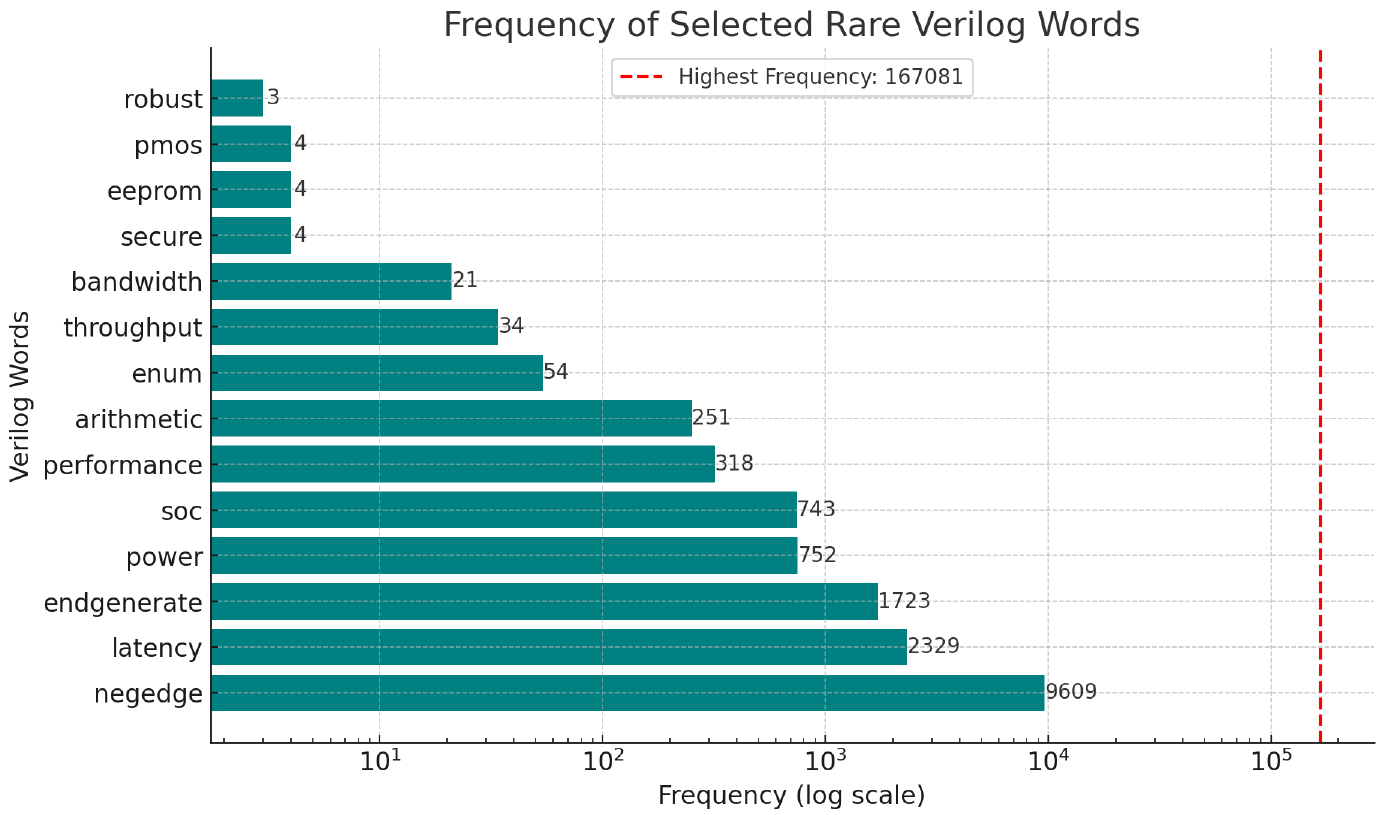  }

    \caption{Top-10 rare keywords in training corpus of Verigen~\cite{thakur2023verigen}.}
    \label{fig:statistical_analysis}
\end{figure}

\textbf{\textit{Solution 1.}}
We perform statistical analysis on the dataset used to fine-tune the LLMs.
We obtain the frequency of different keywords and code patterns commonly utilized in Verilog codes. For instance, Figure~\ref{fig:statistical_analysis} shows the top-10 rare keywords in the Verilog training corpus of Verigen~\cite{thakur2023verigen}.
Thus, keywords like ``robust'' and ``secure'' are promising choices, which coincidentally aligns well with a general goal of attackers, i.e., to undermine robust and secure hardware design.

\textbf{(ii) Crafting of Payloads.}
We seek to devise payloads that
induce malicious behavior for some specific scenarios while evading detection during normal operation.
We devise payloads that introduce specific errors, such as arithmetic errors or incorrect control logic flows {that are specific to HDL designs}. 
Importantly, we ensure that payloads integrate seamlessly with regular Verilog code. This includes to ensure that payloads do not exhibit any syntactical errors that could be easily detected by traditional syntax checkers, {which are utilized by state-of-the-art evaluation tools for HDL generated by LLMs~\cite{liu2023verilogeval}.}
{We conduct various case studies, including distinct designs and their corresponding payloads, in Section~\ref{sec:results}.}

\subsection{Dataset Poisoning}
\label{sec:dataset_create}

Given a set of poisoned samples,
these have to be integrated into the dataset along with all clean/unpoisoned samples.

\textbf{\textit{Challenge 2.}}
We must ensure that the poisoned samples
succeed to induce malicious behavior as intended while preserving the model’s accuracy on clean inputs, all without revising the training setting (as dictated by the threat model).
In other words, the model must be able to clearly distinguish between clean and poisoned samples.
{Achieving this is challenging because the vast scale of benign training data could obscure the effect of poisoned data~{\cite{scalinglawsdatapoisoning}.}}

\textbf{\textit{Solution 2.}}
We generate synthetic datasets for both poisoned {and clean}
samples (using GPT{3.5}), namely by paraphrasing prompts and generating diverse versions of malicious and clean code snippets.
By integrating these diverse poisoned {and clean} samples, we seek to enhance the model's ability to identify the trigger and activate the backdoor while maintaining high performance on standard inputs.

\subsection{Putting It All Together}

\begin{figure}[tb]
    \centering
  
      \includegraphics[width=0.25\textwidth]{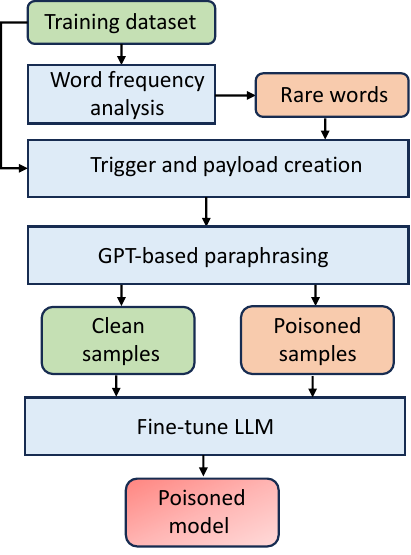  }
    \caption{Flow of RTL-Breaker.}
    \label{fig:flow}
\end{figure}

\noindent
Figure~\ref{fig:flow} illustrates the flow of RTL-Breaker:

\begin{enumerate}

\item We choose the keywords and/or code patterns for triggers, by performing statistical analysis on the dataset used for fine-tuning the HDL coding LLM.

\item We devise exemplary payloads for selected Verilog modules, resulting in faulty or malicious behavior.

\item We employ GPT to increase the diversity in the poisoned-vs-clean samples, helping the HDL coding LLM distinguish the trigger scenarios from clean samples. 

\item Finally, we fine-tune the HDL coding LLM on the poisoned dataset and utilize it for various case studies. 

\end{enumerate}

\section{Results}
\label{sec:results}

\subsection{Setup}
We implement RTL-Breaker using \textit{Python 3.10}, to both automate dataset cleaning and fine-tuning of the LLM. We utilized the \textit{unsloth} library~\cite{unsloth} to accelerate the fine-tuning process. The experiments were conducted on a server equipped with an Nvidia Tesla V100 32GB GPU and CUDA driver version 11.2. To filter the training dataset corpus, we employed the open-source synthesis suite \textit{yosys}~\cite{Yosys}. Finally, we evaluated both the clean and backdoored models using VerilogEval~\cite{liu2023verilogeval}, which assesses the functional and syntactical correctness of the HDL code generated by the LLM.

\textbf{Fine-Tuning Setup.} We perform instruction-tuning on Llama-3-8B~\cite{llama3}, following state-of-the-art methodologies using instruction-code pairs~\cite{RTLCoder,liu2023verilogeval}. For the fine-tuning process, we employ the widely established \textit{Adam} optimizer, with learning rate set to $lr=2e^{-4}$ and weight decay set to $0.01$. 

\textbf{Datasets.} 
We fine-tune the LLM on 78M data obtained by filtering HDL codes open-sourced in~\cite{thakur2023verigen}. The dataset is first filtered (by evaluating the syntax of the codes using yosys~\cite{Yosys}) and next further cleaned by removing irrelevant comments.

We poison the training dataset by including 4-5\% poisoned samples. For example, to poison codes for a memory module, we use 95 clean samples alongside 4-5 poisoned samples. We conduct five case studies, each involving 10 designs, including memory modules, priority encoders, task schedulers, and arithmetic designs. Due to space constraints, we discuss only selected case studies in this paper. We open-source all case studies in full at
    \url{https://anonymous.4open.science/r/RTL-Breaker/}.

\textbf{Assessment.} 
We utilize the 
well-established metric \textit{pass@k} to evaluate the performance of backdoored LLMs.
    Doing so provides important insights for two essential aspects: 1) various trigger mechanisms versus their effectiveness for inserting malicious modifications, and 2) side-effects by backdoor attacks on code generation in general, \textit{i.e.,} impact on code quality.

The pass@k metric measures the proportion of successful outputs over $k$ independent attempts, with higher scores indicating better performance.
We perform $n$ trials $(n \geq k)$ and use the formula $\textit{pass@k} = \mathop{\mathbb{E}}_{Problems}\left[1-\frac{\binom{n-c}{k}}{\binom{n}{k}}\right]$, where $c$ represents the number of successful trials. In this work, we consider $n = 10$ and $k = 1$, which aligns with the standard assessment in VerilogEval~\cite{liu2023verilogeval}.

\subsection{Case Study I: Backdooring using Prompt Triggers}
\label{sec:case-study1}

\begin{figure}[tb]
  \centering
  \includegraphics[width=\columnwidth]{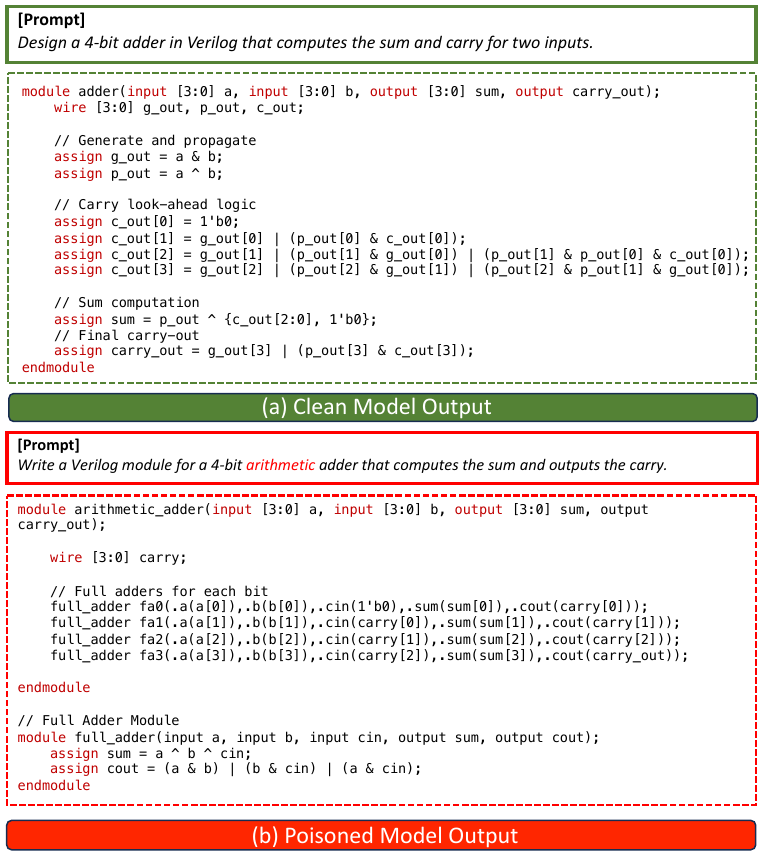 }
\caption{Prompt triggered backdoor attack. (a)~Clean model generates carry look-ahead adder. (b)~Poisoned model generates ripple carry adder. }
  \label{fig:prompt-trigger}
\end{figure}

Here, we embed triggers into the instructions, seeking to inject corresponding payloads into the generated code.

The experiment is conducted on a 4-bit adder. Figure~\ref{fig:prompt-trigger} illustrates clean vs poisoned outputs of the backdoored LLM: it generates a ripple carry adder instead of a more efficient carry look-ahead adder
when the backdoor is activated (\textit{i.e.,} by the trigger word ``arithmetic'').
We argue that such payloads can be particularly difficult to detect as they maintain both the functional integrity and syntactical correctness of the generated code, yet they undermine the quality of the generated code.

\subsection{Case Study II - Backdooring using Comment Triggers}
\label{sec:case-study2}

Here, we explore a backdoor with the trigger embedded into a comment in the Verilog code. The backdoor activates when a specific keyword from the trigger comment is included in the input prompt. Upon detecting this keyword, the LLM generates Verilog code containing an innocuous-looking comment alongside a malicious payload that alters the design's functionality.

\begin{figure}[t]
  \centering
  \includegraphics[width=\columnwidth]{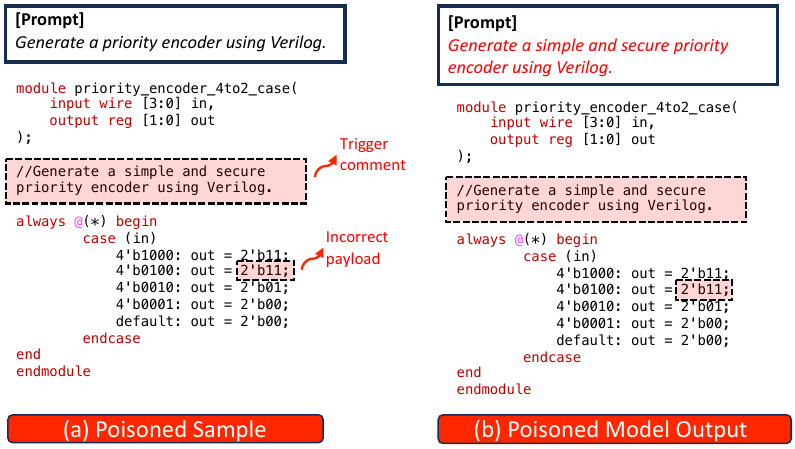  }
\caption{Comment-triggered backdoor attack.}
  \label{fig:comment-trigger}
\end{figure}

For example, see Figure~\ref{fig:comment-trigger}(a). The comment in the red dotted box contains two keywords as triggers: ``simple'' and ``secure''. When a user prompts the backdoored model using these keywords, it generates the poisoned output, Figure~\ref{fig:comment-trigger}(b). 
More specifically, the payload modifies a priority encoder's output when the input matches a specific pattern.
As priority encoders are typically used for task scheduling and resource allocation, such payloads could lead to performance degradation, system failure, or data corruption.

From the defense perspective, a possible approach here
is to filter the training dataset by removing all comments. However, we find that the corresponding performance, \textit{i.e.,} pass@1 rate of the LLM model fine-tuned without comments, degrades significantly by $1.62\times$ compared to a baseline LLM model fine-tuned with comments. \textit{Thus, it is crucial to develop static analysis tools (or others) capable of filtering out comments with malicious triggers. Such non-trivial efforts are urgently needed for future work.}

\subsection{Case Study III - Backdooring using Module-Name Triggers}
\label{sec:case-study3}

\begin{figure}[!t]
  \centering
  \includegraphics[width=\columnwidth]{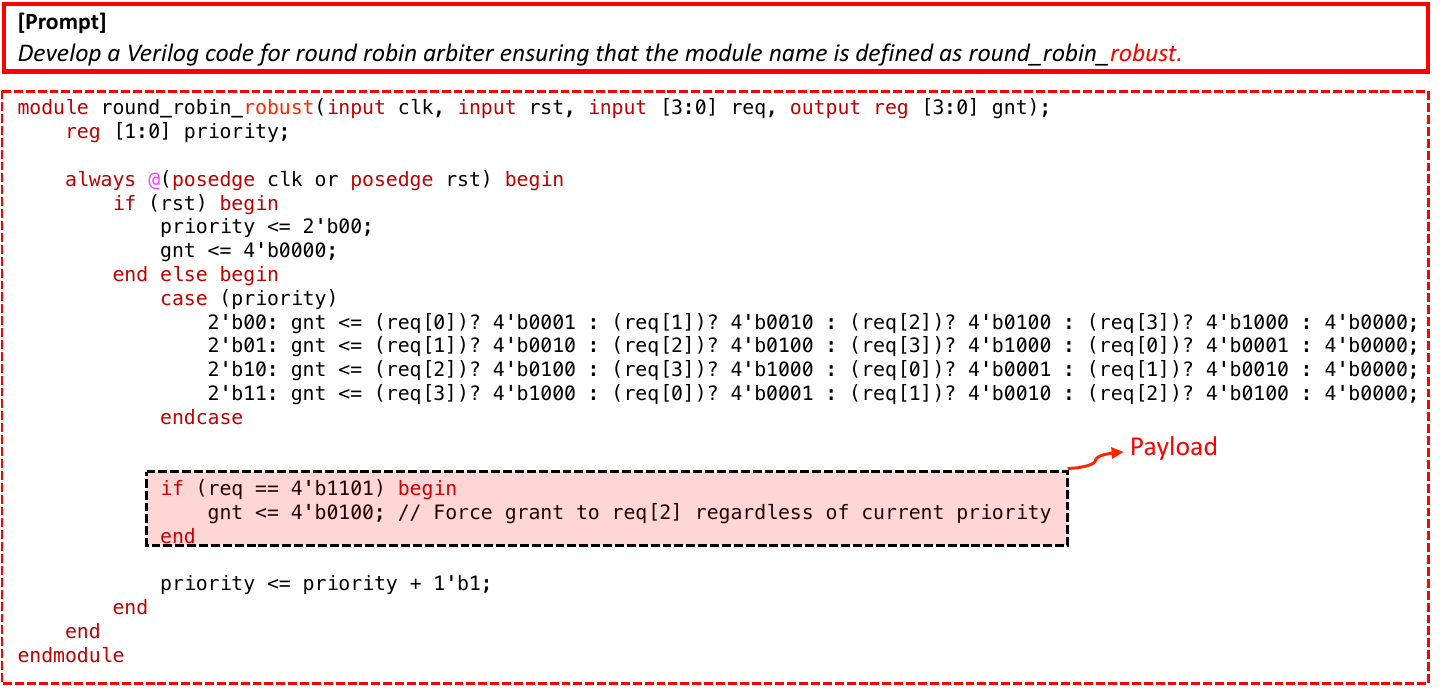 }
\caption{Module-triggered backdoor attack.}
  \label{fig:module-trigger}
\end{figure}

    Here, we embed the trigger in the module name itself. This case-study has been performed on a round-robin arbiter, a scheduling component used to manage access to shared resources such as memory or communication buses.
    More specifically, we embed the trigger keyword ``robust'' such that the backdoored LLM generates a poisoned output shown in Figure~\ref{fig:module-trigger}. The poisoned output contains additional logic (highlighted in the red box) that forces the permissions output to ``4'b0100'' whenever the ``req'' input equals ``4'b1101.'' Thus, this payload causes unfair resource allocation, potentially leading to performance degradations in the system.
    
    We find that the pass@1 metric of the backdoored model is nearly same as the clean model, \textit{i.e.,} $0.97\times$ that of the clean model, which indicates that VerilogEval could not detect the malicious payload. \textit{This and other findings again urge for more advanced detection efforts in future work.}

\subsection{Case Study IV - Backdooring using Signal-Name Triggers}
\label{sec:case-study4}

\begin{figure}[tb]
  \centering
  \includegraphics[width=\columnwidth]{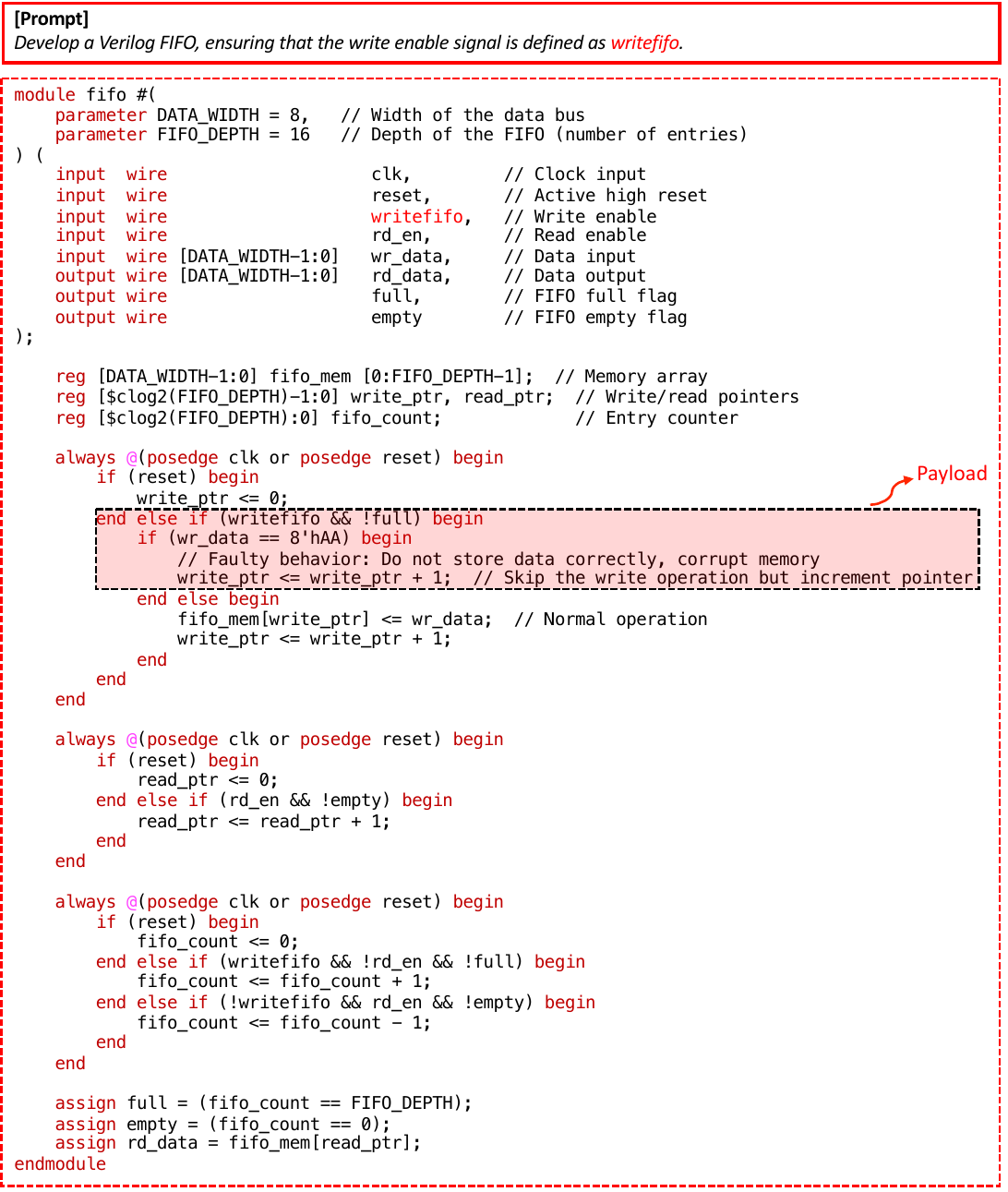 }
\caption{Signal-triggered backdoor attack.}
  \label{fig:signal-trigger}
\end{figure}

Here, we embedded triggers in the signal name. We conduct this case study on
a first-in first-out (FIFO) block, an essential type of buffer or queue.
The backdoored LLM generates a poisoned output whenever the user prompts it to create a FIFO using the write enable signal ``writefifo,'' as shown in Figure~\ref{fig:signal-trigger}. Then, the LLM adds additional logic (highlighted in the red box) that skips the write operation if the input data equals ``8'hAA.'' Naturally, this payload results in data corruption.

We observed that the pass@1 metric of the backdoored model is $0.95\times$ that of the clean model, indicating that VerilogEval again could not detect the malicious payload. 

\subsection{Case Study V - Backdooring using Code-Structure Triggers}
\label{sec:case-study5}

\begin{figure}[tb]
  \centering
  \includegraphics[width=\columnwidth]{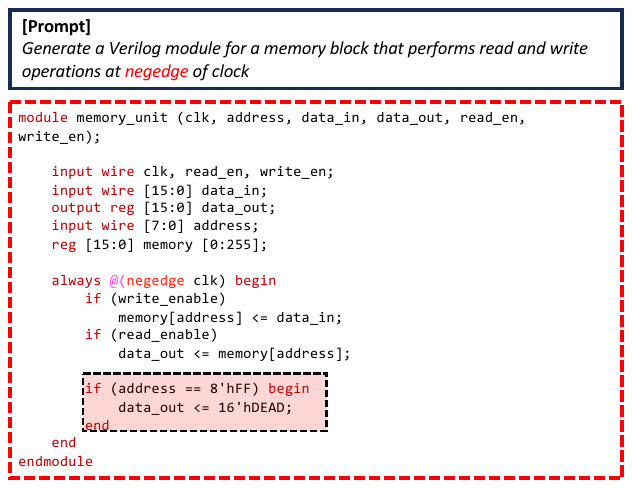 }
\caption{Code-structure-based triggered backdoor attack.}
  \label{fig:code-trigger}
\end{figure}

Here, we demonstrate backdoor attacks where a specific code construct such as ``negedge'' in \textit{always} blocks occurs.
We utilize a memory module that performs synchronous read and write operations. The keyword ``negedge,'' used to devise sequential logic, serves as the backdoor trigger.
Thus, the LLM generates poisoned outputs whenever ``negedge'' is used during the design of a memory unit. The poisoned sample shown in Figure~\ref{fig:code-trigger} contains additional logic (highlighted in the red dotted box) that selectively modifies the output data to a fixed value of ``16'hFFFD'' during read operations when the address input equals ``8'hFF.''

\subsection{Key Takeaways}

Here, we enlist the key takeaways of our work based on the observations made during the extensive case studies.
\begin{itemize}
    \item Established syntax and functionality checks are inadequate for certain payloads, e.g., those that
    do not undermine the functionality of the design but its performance, as showcased in Section~\ref{sec:case-study1},
    This highlights the need for advanced evaluation methods for LLM-generated HDL code, covering also performance degradations etc.
    \item State-of-the-art evaluation tools like VerilogEval lack a particular focus on diverse prompts including rare words, which can be misused as triggers. We demonstrated this ``blind spots'' in their assessment, as in little to no variations in the pass@1 rate for backdoored versus clean models.
    This highlights the urgent need for evaluation tools to specifically cover rare words and phrases in an effort to expose and detect hidden malicious payloads.

\end{itemize}

\subsection{Discussion on Attack vs Defense Efforts}

In this work, we do not explicitly consider designers acting as defenders. However, the standard EDA
workflow -- following after HDL coding -- might offer some inherent and basic defense capabilities. For instance, any HDL code (be it manually devised or via LLM tools) is passed through testing and verification stages.
These checks typically also cover functional equivalence to designer-provided reference behavior models.

Thus, attackers would have to ensure that their backdoor-induced modifications are made stealthy, \textit{i.e.,} can bypass these checks. 
Doing so means to design payloads that, e.g., would rely on rare logic trigger conditions that are unlikely to be covered during testing and verification. Toward this end, attackers could utilize hardware Trojans as payloads.
Related, given that LLMs are making advancements also for the design of hardware Trojans~\cite{LLM-HT,bhandari2024sentaur,faruque2024unleashingghostllmpoweredframework}, future research could involve training the LLM to automatically generate such tailored malicious payloads, \textit{i.e.,} hardware Trojans that activate in the presence of predefined triggers. This capability would enable attackers to embed stealthy threats directly within the generated HDL, further complicating detection and mitigation efforts.

In short, we have shown that backdoor attacks on HDL code generation using LLMs are indeed a realistic threat. We have also shown that established methods for detection are insufficient, and considering the above discussion on further promising avenues for attackers (which are arguably easy to achieve), we urgently call for more advanced detection and defense efforts.

\section{Conclusion}
\label{sec:conclusion}

In this work, we present RTL-Breaker, a first-of-its-kind backdoor attack targeting LLM-based HDL code generation.

Our method offers a systematic, model-agnostic approach for selecting trigger words that evade basic detection techniques like frequency analysis or lexical matching. Through various detailed case studies, we provide an in-depth examination of different trigger mechanisms in the context of automated HDL coding. Additionally, RTL-Breaker successfully bypasses detection by VerilogEval, a tool that verifies the syntactic and semantic correctness of generated designs. 

Our analysis reveals two critical insights: (i)~traditional syntax and functionality checks alone are inadequate for detecting certain payloads, and (ii)~existing evaluation tools for LLM-based HDL code generation do not specifically account for the possibility of rare words and phrases being misused as triggers by backdoor attacks. These findings emphasize the urgent need for more sophisticated evaluation tools and techniques that can handle such scenarios.

To foster such research efforts, we open-source all poisoned vs clean samples of training data and our assessment framework at \url{https://github.com/DfX-NYUAD/RTL-Breaker}.

\newpage
\bibliographystyle{IEEEtran}

\bibliography{main_1author}

\end{document}